\documentclass[3p,twocolumn]{elsarticle}
\usepackage{amstext}
\usepackage{graphicx}% Include figure files
\usepackage{dcolumn}% Align table columns on decimal point
\usepackage{bm}% bold math
\begin{document}

%\preprint{}

\begin{frontmatter}

\title{A reservoir trap for antiprotons}

\author[IRU,CERN]{C. Smorra\corref{cor1}}\ead{christian.smorra@cern.ch}
\author[IRU]{A. Mooser}
\author[IRU,MPIK]{K. Franke}
\author[IRU,TU]{H. Nagahama}
\author[IRU,IPHM]{G. Schneider}
\author[IRU,TU]{T. Higuchi}
\author[RIKEN]{S.V. Gorp}
\author[MPIK]{K. Blaum}
\author[TU]{Y. Matsuda}
\author[GSI]{W. Quint}
\author[IPHM,HIM]{J. Walz}
\author[RIKEN]{Y. Yamazaki}
\author[IRU]{S. Ulmer}

\address[IRU]{RIKEN, Ulmer Initiative Research Unit, 2-1 Hirosawa, Saitama, Japan}
\address[CERN]{CERN, CH-1211 Geneva, Switzerland}
\address[MPIK]{Max-Planck-Institut f\"ur Kernphysik, Saupfercheckweg 1, D-69117 Heidelberg, Germany}
\address[TU]{Graduate School of Arts and Sciences, University of Tokyo, Tokyo 153-8902, Japan}
\address[IPHM]{Institut f\"ur Physik, Johannes Gutenberg-Universit\"at D-55099 Mainz, Germany}
\address[RIKEN]{RIKEN, Atomic Physics Laboratory,  Wako, Saitama 351-0198, Japan}
\address[GSI]{GSI - Helmholtzzentrum f\"ur Schwerionenforschung, D-64291 Darmstadt, Germany}
\address[HIM]{Helmholtz Institut Mainz,  D-55099 Mainz, Germany}

\cortext[cor1]{Corresponding author}

\date{\today}

\begin{abstract}
We have developed techniques to extract arbitrary fractions of antiprotons from an accumulated reservoir, and to inject them into a Penning-trap system for high-precision measurements. In our trap-system antiproton storage times $>1.08$ years are estimated. The device is fail-safe against power-cuts of up to 10$\,$hours. This makes our planned comparisons of the fundamental properties of protons and antiprotons independent from accelerator cycles, and will enable us to perform experiments during long accelerator shutdown periods when background magnetic noise is low. The demonstrated scheme has the potential to be applied in many other precision Penning trap experiments dealing with exotic particles.
\end{abstract}


\end{frontmatter}

%\maketitle

\begin{figure*}[ht]
        \centerline{\includegraphics[width=18cm,keepaspectratio]{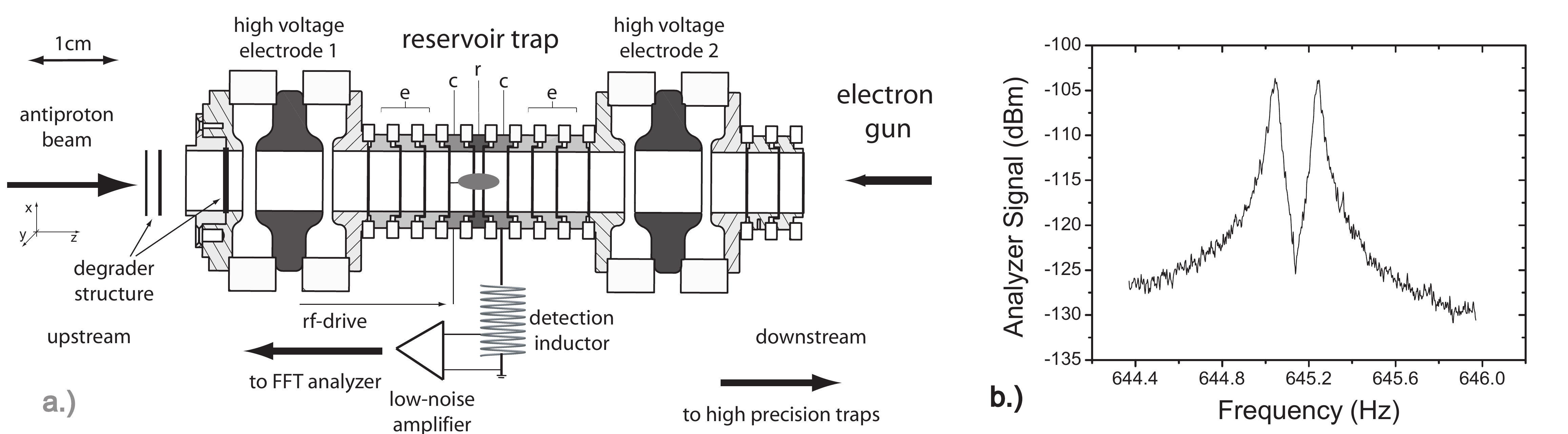}}
            \caption[FEP]{a.) Schematic of the reservoir trap and the crucial ingredients. The trap is in 5-electrode orthogonal design and has an inner diameter of 9$\,$mm. Endcaps, correction electrodes, and the central ring electrode are marked, e, c, and r, respectively. Radio-frequency drives for particle manipulation are applied to the segmented correction electrode placed upstream. The trap is placed between two high-voltage electrodes for antiproton catching. On the upstream side a degrader structure is mounted, downstream an electron gun is installed. The sensitive detection system is connected to an electrode adjacent to the central ring. b.) Noise-dip signal of about 100 captured antiprotons at $T_z$=$5.3(1.1)\,$K. }
            \label{fig:Trap}
\end{figure*}

\emph{Introduction} - Experiments with exotic particles, like antiprotons, positrons, radioactive or highly-charged ions, in Penning traps provide high-precision data for stringent tests of the fundamental laws of nature. For example measurements of the properties of positrons \cite{vandyckjr1987nhp} and antiprotons \cite{JerryAntiproton}, and comparisons to their matter equivalents provide the most stringent tests of $CPT$ symmetry with leptons and baryons \cite{Bluhm1}. Precise tests of bound-state quantum electrodynamics, on the other hand, are based on measurements of the $g$-factor of the electron bound to hydrogen-like ions \cite{Hartmut,Verdu,Sven}. Further experiments are in preparation to enhance the sensitivity of this type of test by performing similar measurements in even stronger fields using $^{208}$Pb$^{81+}$, $^{209}$Bi$^{82+}$ and $^{238}$U$^{91+}$ \cite{WQ}.
Our initiatives at the Baryon Antibaryon Symmetry Experiment (BASE) \cite{UlmerPRL,CCR,Smorra} target both: an improved precision in the proton-to-antiproton charge-to-mass ratio, as well as magnetic moment comparisons of the proton and the antiproton \cite{MooserNature}. Both efforts will provide stringent tests of CPT-invariance using baryons. \\
In most of the above cases the exotic particles are provided by external sources such as accelerators \cite{AD1,HITRAP} or powerful electron beam ion traps \cite{EBIT}. Thus, the experiments depend on accelerator run times and scheduled beamtimes, and data-acquisition time is usually limited. Moreover, the high-power sources produce considerable electric and magnetic noise, which potentially limits experimental precision \cite{JerryAntiproton}. In our case, CERN's antiproton decelerator (AD) storage ring causes background magnetic field fluctuations of about 100$\,$nT during one accelerator cycle, which is on the order of several 10$\,$ppb with respect to the field strengths of typical superconducting solenoids. Thus, it is desirable to perform these kind of high-precision measurements during accelerator shut-down.  \\

\emph{In this article} we report on the design and commissioning of a device to address these challenges - a reservoir trap for exotic particles. The trap is loaded from the AD of CERN. We have developed techniques for lossless extraction of arbitrary fractions of antiprotons from this reservoir, and injection into our precision Penning traps. Thus, we can replace particles in the precision trap cycle by extracting another particle from the reservoir. By detecting clouds of antiprotons for several weeks we obtain estimated antiproton storage times $>1.08\,$years. Together with the developed extraction schemes the long storage time enables operation of the experiment independent from accelerator cycles. The device is fail-safe against power cuts of up to 10$\,$h. Our reservoir trap technique has the potential to be implemented into a manifold of other high-precision Penning-trap experiments which investigate the fundamental properties of exotic particles. This will allow further enhancement of experimental precision in this type of experiments. \\

\emph{Experimental setup} - Our experiment consists of an advanced four-Penning trap system mounted in the horizontal bore of a superconducting magnet at $B_0=1.946\,$T. Of interest for this work is the reservoir trap, shown in Fig$.\,$\ref{fig:Trap} a$.$). The traps consist of five cylindrical oxygen-free-electrolytic (OFE) copper electrodes with an inner diameter of 9$\,$mm, which are in orthogonal and compensated design \cite{gabrielse1989oep}. Upstream and downstream of the trap high-voltage electrodes are placed, which allow application of catching pulses of up to 5$\,$kV. Downstream of the reservoir trap the high-precision Penning traps (not shown in the figure) are located. Transport electrodes connect the reservoir to these traps. Voltage ramps applied to the transport electrodes allow adiabatic particle shuttling along the trap axis. To manipulate the trapped particles with external drives
a radio-frequency line is connected to a segmented correction electrode. This allows for application of dipolar axial excitation signals as well as quadrupolar axial-to-radial coupling drives. The entire trap-assembly is placed between a degrader system with 224$\,\mu$m aluminium equivalent thickness \cite{Degrader} on the upstream side, and a field emission electron-gun on the downstream side. The latter provides electrons for sympathetic cooling of antiprotons \cite{ElectronCooling}. \\
This setup is mounted in a cylindrical OFE trap-chamber with a volume of about 1.2$\,$l, which is closed on both ends with indium-sealed OFE-flanges. A stainless steel vacuum window with a diameter of 9$\,$mm and 20$\,\mu$m thickness is hard soldered to the flange on the upstream side. This window holds the vacuum but is transparent with respect to the 5.3$\,$MeV antiprotons provided by the AD. Before installation, the trap-chamber is pumped through a copper tube which is pinched-off once a pressure $<10^{-6}$ mbar has been reached. By placing the chamber in a $10^{-9}$ mbar insulation vacuum and cooling it to liquid helium temperature, the interior of the trap chamber forms a completely-sealed cryopumped vacuum system. This allows to achieve ultra-low pressure \cite{Thompson}.    \\

\emph{Detection and cooling} - A highly sensitive superconducting single-particle detection system \cite{ulmer2009quality} based on low-noise GaAs transistors and a toroidal NbTi inductor \cite{MooserPRL} is connected to one of the correction electrodes of the trap. Together with the trap capacitance the inductor forms a tuned circuit with resonance frequency $\nu_r=645\,260\,$Hz. At resonance it appears as an effective parallel resistance $R_p=2\pi\nu_r L Q$, where $Q=11\,300$ is the quality factor. With the inductance of 1.72$\,$mH we achieve $R_p=78\,$M$\Omega$. This enables non-destructive measurements of the particle's axial oscillation frequency $\nu_z$ by image current detection \cite{Wine}. Trapped antiprotons are tuned to resonance with the detector by adjusting the trapping voltage. When tuned to resonance the particles are cooled resistively with cooling time constant $\tau_z=(m/R_p)(D/q)^2$, where $m$ and $q$ are the particle's mass and charge, respectively. $D$ is a trap specific length, in our case 10$\,$mm. The modified cyclotron mode at $\nu_+\approx29.65\,$MHz and the magnetron mode at $\nu_-\approx7.02\,$kHz are cooled by application of sideband cooling \cite{Cornell}. Radial mode-temperatures of $T_+=\nu_+/\nu_z\cdot T_z\approx250\,$K and $T_-=\nu_-/\nu_z\cdot T_z\approx60\,$mK are achieved \cite{brown1986gtp}, where the definition $T_k=E_k/k_\text{B}$ with mode energy $E_k$ and $k\in\{+,-\}$ was used. The temperature of the detection system, and thus the temperature of the axial mode is at $T_z=5.3(1.1)\,$K.
\\

\emph{Antiproton loading and cleaning} - In each AD antiproton-shot about 30$\cdot$10$^6$ antiparticles are delivered to our apparatus. A 10$^{-4}$-fraction of the incident particles is degraded to energies below 1$\,$keV. These particles are captured by applying a static high-voltage to high-voltage electrode 2 and a fast high-voltage pulse to high-voltage electrode 1. About $15\,000$ electrons which are loaded before the injection cool the captured antiprotons within a typical interaction time of 10$\,$s from keV- to thermal energies. Next, a strong dipolar rf-drive at the axial frequency of the electrons is applied to remove them from the trap. Afterwards quadrupolar drives at $\nu_z+\nu_-$ and $\nu_+-\nu_z$ are used to sideband-cool the magnetron and the cyclotron modes of the antiprotons, respectively. Subsequently, white noise with a bandwidth of 20$\,$kHz to 500$\,$kHz is applied to the drive electrode, and the trap voltage is lowered to typically 500$\,$mV. This drive evaporates heavy negative ions from the trap. To remove electrons on large radii a 500$\,$ns kick-out pulse opens the trapping potential. Electrons escape from the trap, while the 1836 times heavier antiprotons are not affected by the short pulse. In a last step the magnetron motion of the antiprotons is centered again by application of sideband-coupling. Then, the electron drive is turned on again and the trap potential is simultaneously swept to 300$\,$mV. The entire procedure is repeated multiple times leaving a clean cloud of antiprotons in the trap. \\
To count the number of trapped antiprotons the axial detection system is used. A fast Fourier transform (FFT) of the detector's output signal shows a noise resonance which is caused by thermal Johnson noise $e_n=\sqrt{4k_B T_z \text{Re}(Z(\nu))}$, where $Z(\nu)$ is the impedance of the detector \cite{JohnsonNoise}, and $k_B$ the Boltzmann constant.  Antiprotons cooled to thermal equilibrium with the detection system short this noise $e_n$, and a dip appears on the noise resonance, as shown in Fig$.\,$\ref{fig:Trap} b$.$).
The width of the dip $\Delta\nu_z$ is proportional to the number $N$ of trapped antiprotons \cite{Wine}
\begin{eqnarray}
\Delta\nu_z=\frac{N}{2\pi\tau_z}\,,
\end{eqnarray}
where our measured single particle line-width is at 1.9\,Hz.
By using the procedure described above, we typically prepare 100 to 300 antiprotons per AD-shot. \\

\begin{figure}[htb]
        \centerline{\includegraphics[width=9cm,keepaspectratio]{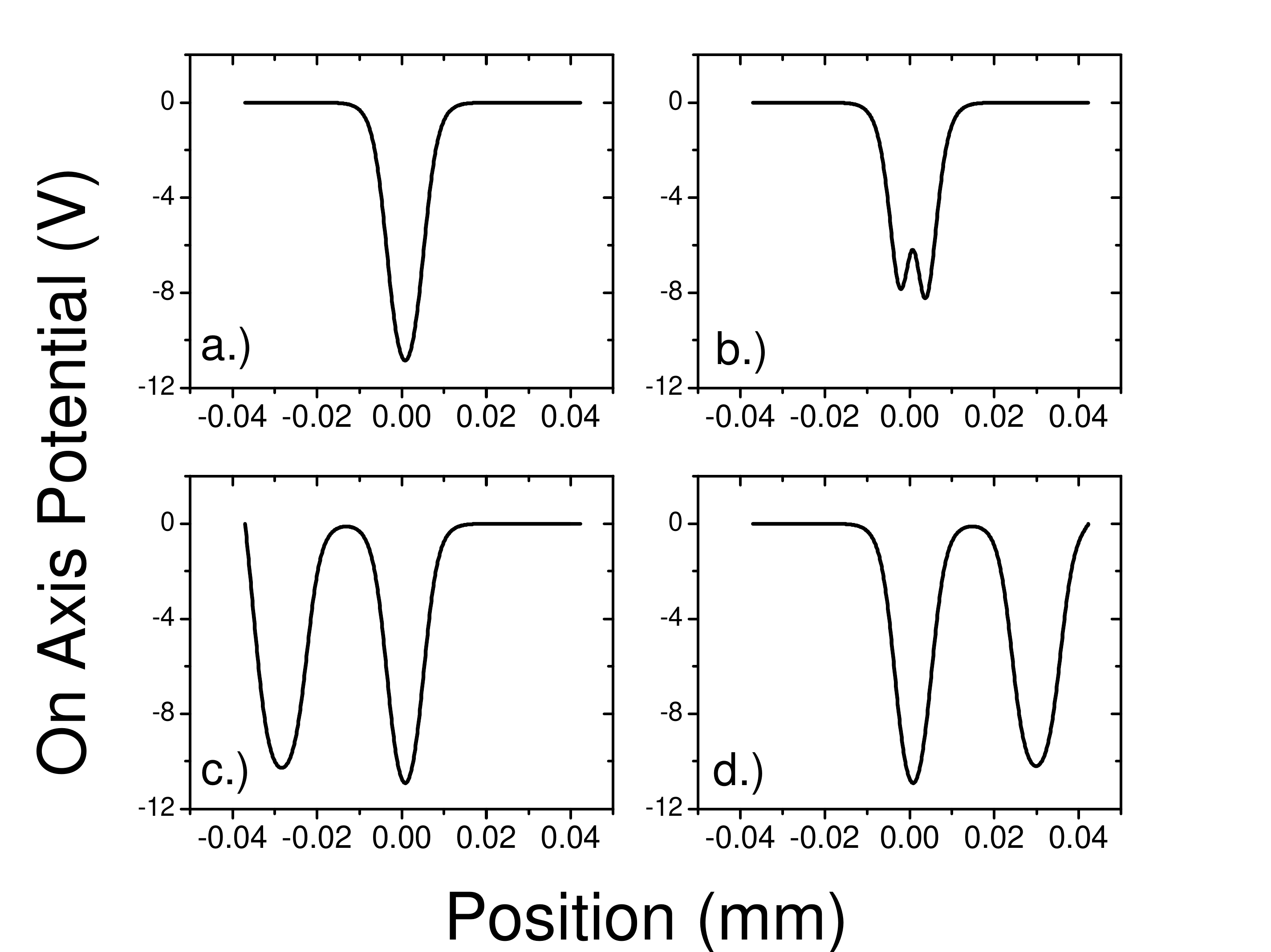}}
            \caption[FEP]{On axis potentials used for particle extraction. a.) Initially applied potential. b.) Potential after separation ramp (see text). c.) Potential after shuttling to high-voltage electrode 1. d.) Potential after shuttling to high-voltage electrode 2.}
            \label{fig:Potentials}
\end{figure}

\emph{Particle extraction from the reservoir} - To extract a certain fraction of antiprotons from the reservoir adiabatic potential ramps are utilized, as shown in Fig$.\,$\ref{fig:Potentials}.
First, the particle cloud is centered into the reservoir trap and the trapping voltage is tuned to 13.5$\,$V (see Fig$.\,$1). Afterwards, a constant electric field $E=0.32\,$V/m$\,\cdot\Delta$V/V is superimposed on the trap, where $\Delta V$ is a potential offset deliberately applied to one of the correction electrodes. This electric field shifts the center of mass of the axial oscillation with respect to the trap center (Fig$.\,$\ref{fig:Potentials} a$.$)). In a next step, the voltage of the central ring electrode is swept from 13.5$\,$V to -13.5$\,$V (Fig$.\,$\ref{fig:Potentials} b$.$)). This separates the particle cloud into two fractions, $F_1$ and $F_2$, respectively. $F_1$ is shuttled to high-voltage electrode 1, while $F_2$ is kept in the trap and the number of particles is analyzed (Fig$.\,$\ref{fig:Potentials} c$.$)). Afterwards, $F_2$ is transported to high-voltage electrode 2 while $F_1$ is simultaneously moved to the trap center and its content is determined as well (Fig$.\,$\ref{fig:Potentials} d$.$)). Including shuttling and number-analysis of both fractions the entire procedure takes in total 120$\,$s.\\

\begin{figure}[htb]
        \centerline{\includegraphics[width=9cm,keepaspectratio]{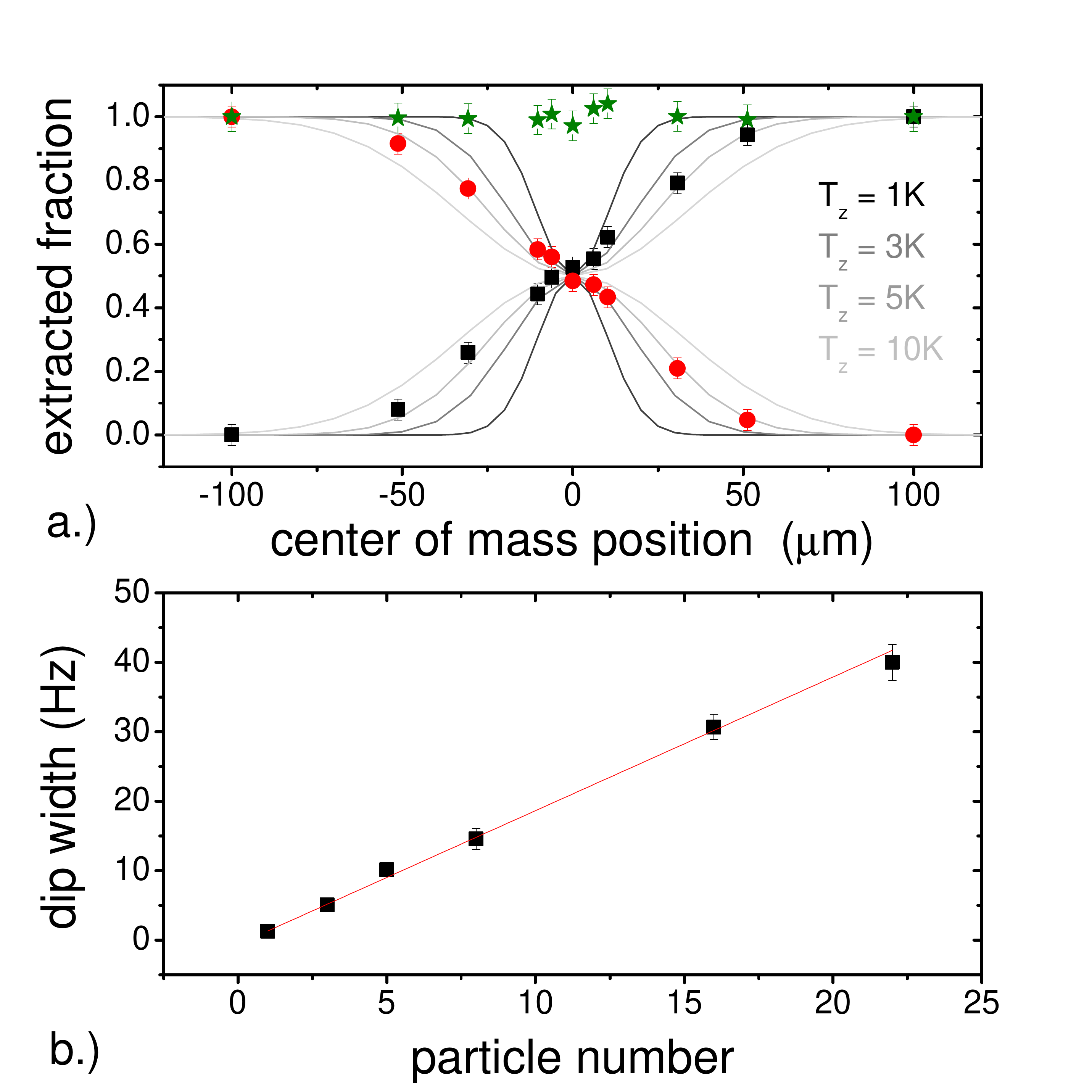}}
            \caption[FEP]{a.) Extracted fraction as a function of the center-of-mass position defined by electric field along the trap axis. The red circles represent particles which were extracted to high-voltage electrode $1$, the black squares represent the fraction which was first kept in the trap and subsequently shuttled to high-voltage electrode $2$. The solid lines represent results of calculations with the axial temperature as array-parameter.  b.) Width of the dip as a function of extracted particles.}
            \label{fig:Separation}
\end{figure}

\emph{Results} of this extraction procedure are shown in Fig$.\,$\ref{fig:Separation} a$.$). The abscissa represents the center of mass positions of the particles due to different electric fields superimposed on the trap. The red circles and the black squares represent the fractions $F_1$ and $F_2$, respectively. The solid lines are from analytical calculations using 1-dimensional Boltzmann-statistics $w(E)=k_BT_z\exp\left(-E/(k_BT_z)\right)$ and integrating
\begin{eqnarray}
N_{down}=C\cdot\int_{z_0}^{\infty} dz z \exp\left(-\frac{2\pi^2 m(\nu_z z)^2}{k_B T_z}\right)\, ,
\end{eqnarray}
with $N=N_\text{up}+N_\text{down}$, where $C$ is a normalization constant. Here, $N_{up}$ is the number of particles separated to high-voltage electrode 1. The measured data are in perfect agreement with the independently measured axial mode-temperature of 5.3(1.1)$\,$K.

This data-set was taken by using antiprotons from a single AD-shot. After each individual separation cycle the particle clouds were merged by reversing the separation sequence. The green stars show the sum of both extracted fractions normalized to the number $N_0$ of antiprotons counted before the first separation cycle. This indicates, that during the entire measuring procedure the particle number was constant and the applied operations are lossless within the uncertainties. \\
Figure \ref{fig:Separation} b$.$) shows absolute numbers of antiprotons which were extracted from a reservoir of about 100 particles. By applying the above scheme we were able to consistently extract arbitrary fractions of particles, in this experiment starting from 22(1) antiprotons down to a single one. For single particle extraction we typically separate fractions $F_1\gg F_2$, and apply a slightly modified version of the above procedure to $F_2$, using the trap and high-voltage electrode 2. Once a single particle has been prepared, it was injected into the high-precision traps, while the remaining cloud was kept in the reservoir. Particles lost in the precision measurement cycle were replaced by extracting and suspending another particle to our precision trap system. %By utilizing this scheme we conducted with one AD antiproton-shot experiments for five weeks.
In our previous proton measurements \cite{UlmerPRL2} we destructively prepared single particles from a cloud of trapped protons by heating and evaporation. It took typically 1$\,$h-2$\,$h to prepare a single proton in the trap. In our antiproton apparatus particle extraction from the reservoir is lossless, and once a clean cloud of antiprotons has been prepared, extraction of a single particle takes only a few extraction trials of 120$\,$s each. \\

\emph{Storage time in the trap} - By accumulating all data recorded within an experiment time of 3 months, where different particle numbers were prepared and stored in the reservoir, we did not observe any antiproton loss. The total accumulated trapping time of an equivalent single particle is $t_o>1.56\,$yrs. From this we extract an integrated antiproton storage time of $t_{1/2}=\text{ln}{2}\cdot t_o > 1.08\,$yrs. To secure the trapping system from particle loss due to power cuts we operate DC-supplies, control-PC, and radio-frequency instruments on uninterruptable power supplies. Operation of the trap during power-cuts of up to 10$\,$h was demonstrated experimentally. \\

\emph{Conclusion - }We have developed a reservoir trap for antiprotons. Experimental routines to extract single particles from this reservoir, and to inject them into an adjacent high-precision Penning-trap system were established and described in the article.
The trap allows us to perform experiments independent from accelerator cycles, and to run experiments during accelerator-shutdown. This technique will be applied in our planned experiments to compare the fundamental properties of protons and antiprotons with high precision \cite{Smorra,Degrader}. Moreover, it has the potential to be applied in other high-precision Penning-trap experiments dealing with exotic particles as in \cite{WQ,HITRAP}.
\\

\emph{We acknowledge} support from the AD group of CERN. We would like to express our strong gratitude towards L. Bojtar and F. Butin. Financial support of RIKEN Initiative Research Unit Program, RIKEN President Funding, RIKEN Pioneering Project Funding, the Max-Planck Society, the IMPRS-PTFS, the CERN fellowship program, the BMBF, the Helmholtz-Gemeinschaft, and the EU (ERC Advanced Grant No. 290870-MEFUCO) is acknowledged.

\end{document}